\documentclass[prl,showpacs,twocolumn,superscriptaddress]{revtex4-1}

\usepackage{amsfonts,amsmath,amssymb,graphicx}
\usepackage[usenames,dvipsnames]{xcolor}
\usepackage{pdfpages}

\usepackage{color,colordvi}
\definecolor{blue}{rgb}{0,0,1}

\newcommand{\Exp}[1]{e^{#1}}
\newcommand{\ket}[1]{|#1\rangle}
\newcommand{\moy}[1]{\langle#1\rangle}
\newcommand{\phiqb}{\varphi_\mathrm{qb}}

\newcommand{\In}{\mathrm{in}}
\newcommand{\out}{\mathrm{out}}
\newcommand{\no}{\bar{n}_0}
\newcommand{\SNR}{\mathrm{SNR}}
\newcommand{\erfc}{\mathrm{erfc}}

\begin{document}

\title{Heisenberg-limited qubit readout with two-mode squeezed light}

\author{Nicolas Didier}
\affiliation{Department of Physics, McGill University, 3600 rue University, Montreal, Quebec H3A 2T8, Canada}
\affiliation{D\'epartment de Physique, Universit\'e de Sherbrooke, 2500 boulevard de l'Universit\'e, Sherbrooke, Qu\'ebec J1K 2R1, Canada}
\author{Archana Kamal}
\affiliation{Research Laboratory of Electronics, Massachusetts Institute of Technology, Cambridge, Massachusetts 02139, USA}
\author{William D. Oliver}
\affiliation{Research Laboratory of Electronics, Massachusetts Institute of Technology, Cambridge, Massachusetts 02139, USA}
\affiliation{MIT Lincoln Laboratory, 244 Wood Street, Lexington, Massachusetts 02420, USA}
\author{Alexandre Blais}
\affiliation{D\'epartment de Physique, Universit\'e de Sherbrooke, 2500 boulevard de l'Universit\'e, Sherbrooke, Qu\'ebec J1K 2R1, Canada}
\affiliation{Canadian Institute for Advanced Research, Toronto, Ontario M5G 1Z8, Canada}
\author{Aashish A. Clerk}
\affiliation{Department of Physics, McGill University, 3600 rue University, Montreal, Quebec H3A 2T8, Canada}

\begin{abstract}
We show how to use two-mode squeezed light to exponentially enhance cavity-based dispersive qubit measurement.  Our scheme enables true Heisenberg-limited scaling of the measurement,
and crucially, is not restricted to small dispersive couplings or unrealistically long measurement times.  It involves coupling a qubit dispersively to two cavities, and making use of a symmetry in the dynamics
of joint cavity quadratures (a so-called quantum-mechanics-free subsystem).  We discuss the basic scaling of the scheme and its robustness against imperfections, as well as a realistic implementation in 
circuit quantum electrodynamics.  
\end{abstract}


\maketitle

{\it Introduction.-- }
Research in quantum metrology has established that squeezed light and entanglement are key resources needed to approach truly fundamental quantum bounds on measurement sensitivity~\cite{Giovannetti_2004}.  Perhaps the best known application is interferometry: by injecting squeezed light into the dark port of an interferometer, one dramatically enhances its sensitivity to small phase shifts \cite{Caves_1981,Yurke_1986}, reducing the imprecision below the shot-noise limit.
Many of these ideas for squeezing-enhanced measurement were first motivated by gravitational wave detection \cite{Braginsky_1980,Caves_1980,Braginsky-Khalili-Thorne}, and have recently been implemented in current-generation detectors \cite{LIGO_2011,LIGO_2013}.  More generally, squeezed light has been used to enhance the measurement sensitivity in optomechanics~\cite{Iwasawa_2013} and even biology~\cite{Taylor_2013}.

\begin{figure}
\includegraphics[width=\columnwidth]{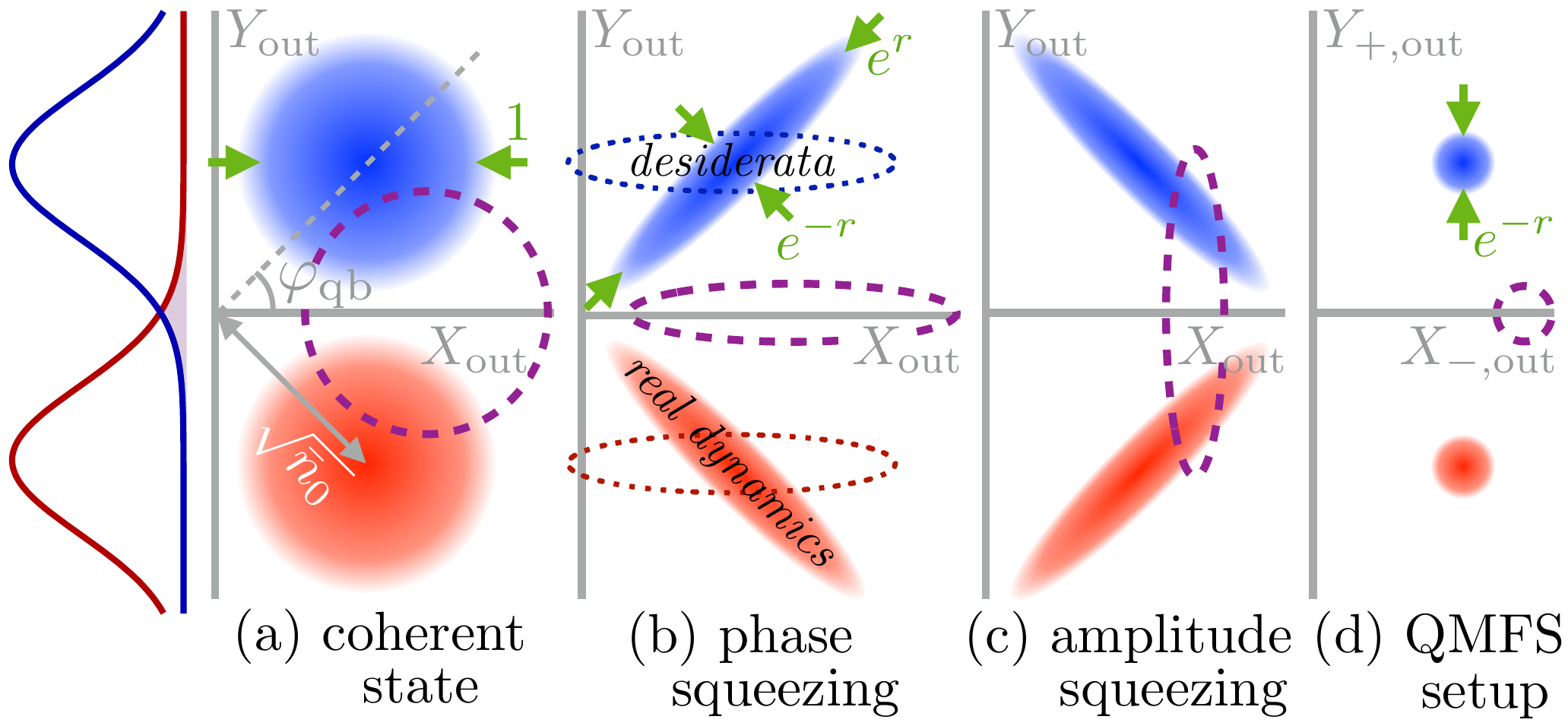}
\caption{Phase-space representation of dispersive qubit readout for different input states:  
(a)~coherent state, 
(b)~single-mode phase-squeezed state, 
(c)~amplitude-squeezed state, 
(d)~two-mode squeezed state in the QMFS $(X_-,Y_+)$.
The purple dashed lines represent the input state and the blobs represent the output fields. 
The input state is displaced along the $X$-axis and the signal is encoded in the quadrature corresponding to the $Y$-axis with homodyne detection; 
as depicted in the leftmost panel, the readout error corresponds to the overlap of the two marginals. 
Dispersive interaction with the qubit rotates the output field by the angle $\phiqb$ for the ground state $\ket{0}$ (in blue) and $-\phiqb$ for the excited state $\ket{1}$ (in red).
Ideally, one wants the output state to be phase squeezed regardless of qubit state [dotted ``desiderata'' states in (b)]; this is not possible when using single-mode squeezing due to the qubit-induced rotation.  Our new QMFS scheme [panel (d)] does not suffer from this problem.}
\label{figscheme}
\end{figure}

Ultrasensitive detection is also essential for quantum information processing where fast, high-fidelity qubit readout is required to achieve fault-tolerant quantum computation~\cite{Nielsen-Chuang}.  A ubiquitous yet powerful approach is dispersive readout, where a qubit couples to a cavity such that the cavity frequency depends on the qubit state, see e.g.~\cite{Blais_2004}.  The readout consists in driving the initially empty cavity with a coherent tone, resulting in a qubit-state dependent cavity field which is displaced in phase space from the origin [see Fig.~\ref{figscheme}(a)].  
High-fidelity readout can then be obtained by measuring the output field quadratures. 
This is the standard approach used in state-of-the-art experiments with superconducting qubits, e.g.~\cite{Jeffrey_2014,Liu_2014,Zin_2013}.

As with interferometry, one might expect that dispersive qubit measurement could be  enhanced by using squeezed light. 
The most obvious approach would be to squeeze the phase quadrature of the incident light [i.e.~$Y$ in  Fig.~\ref{figscheme}(b)], thus reducing the overlap between the two pointer states.  
As discussed recently in Ref.~\onlinecite{Barzanjeh_2014} the situation is not so simple, as the dispersive interaction will  lead to a qubit-dependent rotation of the squeezing axis. 
Unlike standard interferometry, this rotation is a problem, as optimal dispersive qubit readout involves large couplings and hence large rotations.  
Further complexity arises from the fact that this rotation is frequency dependent.  
The upshot is that measurement always sees the amplified noise associated with the antisqueezed quadrature of the incident light, limiting the fidelity improvement from using squeezing to modest values and preventing true Heisenberg scaling \cite{Barzanjeh_2014}. 

\begin{figure}
\includegraphics[width = 0.7\columnwidth]{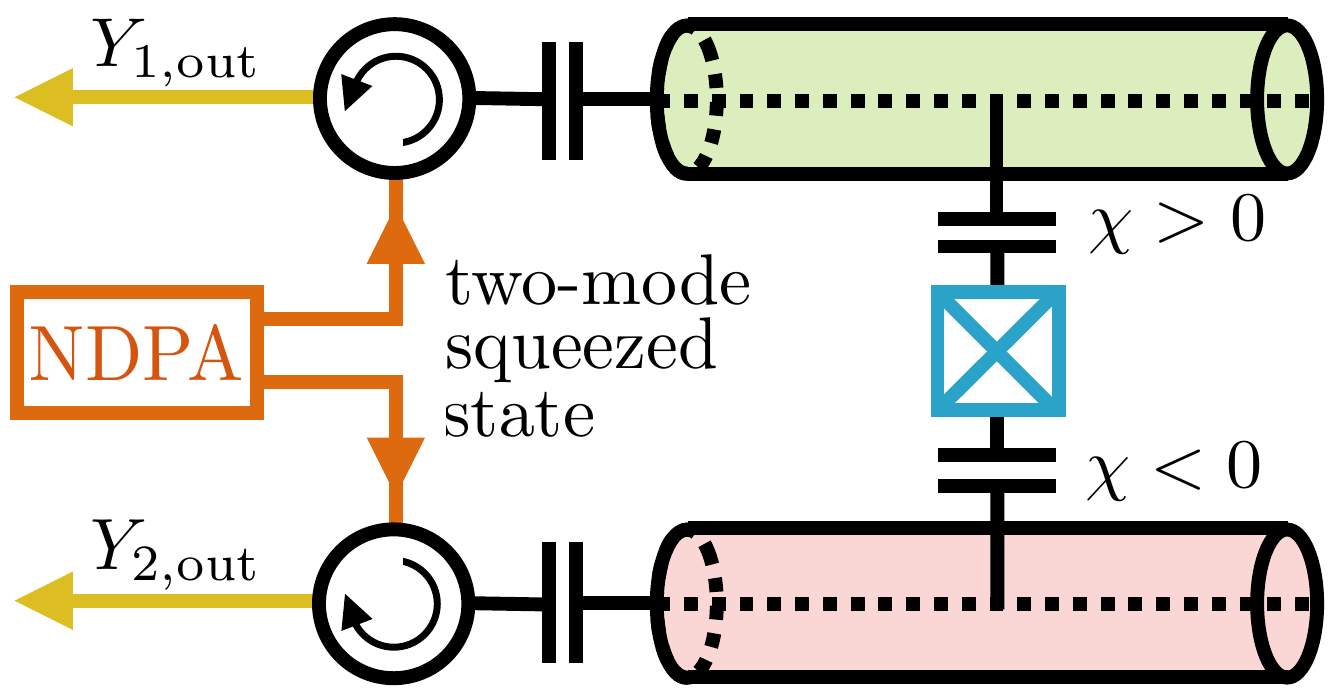}
\caption{Circuit QED implementation of QMFS dispersive qubit readout with two-mode squeezed states
produced by a nondegenerate paramp (NDPA).
The signal is encoded in the joint quadrature $Y_{+,\out}\propto Y_{1,\out}+Y_{2,\out}$; see text for details.}
\label{figcircuit}
\end{figure}

Despite the above difficulties, we show in this Letter that it is indeed possible to substantially improve dispersive qubit measurements using squeezed input states.  
Our proposed scheme involves using two-mode squeezed states in a two-cavity-plus-qubit system (see Fig.~\ref{figcircuit}), 
which can lead to exponential enhancement of the signal-to-noise ratio (SNR) in dispersive measurement and achieves true Heisenberg-limited scaling.  
This is possible even
for large qubit-induced phase shifts, and is thus in stark contrast to previous schemes using two-mode squeezing for interferometry \cite{Yurke_1986} or qubit readout~\cite{Barzanjeh_2014}.  

The key to our scheme is the use of a special dynamical symmetry, whereby 
two commuting collective quadratures exhibit a simple rotation as a function of time.  
As these quadratures commute, they constitute a so-called ``quantum-mechanics-free subsystem'' (QMFS)~\cite{Tsang_2012} and can both be simultaneously squeezed.  
The upshot is that one can effectively make a dispersive qubit measurement where now the uncertainties associated with the two pointer states are not limited by the uncertainty principle [see Fig.~\ref{figscheme}(d)]. Though the scheme is extremely general, for concreteness we explicitly discuss an implementation in circuit quantum electrodynamics (QED) using a transmon qubit~\cite{Koch_2007}, as depicted in Fig.~\ref{figcircuit}. 

The dynamical symmetry used in our two-mode scheme crucially relies on one of the cavities having an effective negative frequency; it is thus
related to an idea first discussed in the context of measurement by Tsang and Caves~\cite{Tsang_2010} and Wasilewski \textit{et al.}~\cite{Wasilewski_2010}, and which has since been applied to other systems \cite{Tsang_2012,Woolley_2013,Meystre_2013}.  While many applications use the idea to suppress the effects of backaction \cite{Wasilewski_2010,Woolley_2013,Meystre_2013}, we instead use it as an effective means to exploit squeezed input light.  Unlike previous studies, we calculate here the scaling of the resulting measurement sensitivity, showing that one obtains Heisenberg-limited scaling with incident photon number. 

{\it Dispersive measurement and standard squeezing.-- }We start by reviewing the simplest setup 
where a qubit dispersively couples to a single-sided cavity (frequency $\omega_1$) with the Hamiltonian 
$H=\left( \omega_1 + \chi \hat{\sigma}_z \right)\hat{a}^\dag\hat{a}$~\cite{Blais_2004}.
Standard dispersive readout involves driving the input port of the cavity with a coherent tone at the cavity frequency (photon flux $\no \kappa/4$, $\kappa$ is the cavity damping rate).  
As illustrated in Fig.~\ref{figscheme}(a), as a consequence of the dispersive coupling, the output field
is rotated by the angle $\phiqb=2\arctan(2\chi/\kappa)$ if the qubit is in the ground state $\ket{0}$ and by $-\phiqb$ for the excited state $\ket{1}$. 
Writing the output field as 
$\hat{a}_{\rm out}(t) = e^{-i \omega_1 t}(\hat{X}_{\rm out} + i \hat{Y}_{\rm out})/2$, for a displacement along the real axis $X_{\rm out}$, the signal of the qubit state 
is  encoded in the phase quadrature $Y_{\rm out}$; this quadrature is then recorded with homodyne detection. 

Measuring $Y_{\rm out}$ for an integration time $\tau$ corresponds to evaluating the dimensionless measurement operator 
$\hat{M}=\sqrt{\kappa}\int_0^\tau\mathrm{d}t\hat{Y}_\out(t)$. 
The signal is the qubit-state dependent expectation value $M_S=\moy{\hat{M}}$ and is the same for all the injected states depicted in Fig.~\ref{figscheme}.
The imprecision noise is the variance of the noise operator $\hat{M}_N=\hat{M}-M_S$. 
The signal-to-noise ratio 
$\SNR \equiv \left|M_{S,\ket{0}}-M_{S,\ket{1}}\right|/(\moy{\hat{M}_{N,\ket{0}}^2}+\moy{\hat{M}_{N,\ket{1}}^2})^{1/2}$
is, for this coherent state dispersive readout,
$\SNR_\alpha(\tau)\simeq |\sin\phiqb|\sqrt{2\no\kappa\tau}$~\cite{Gambetta_2008,Clerk_2010}.
As expected, the SNR is maximized for a phase $\phiqb = \pi/2$; it also scales as $\sqrt{ \no}$, akin to standard quantum-limit scaling in interferometry~\cite{Giovannetti_2004}.

\begin{figure}
\includegraphics[width=\columnwidth]{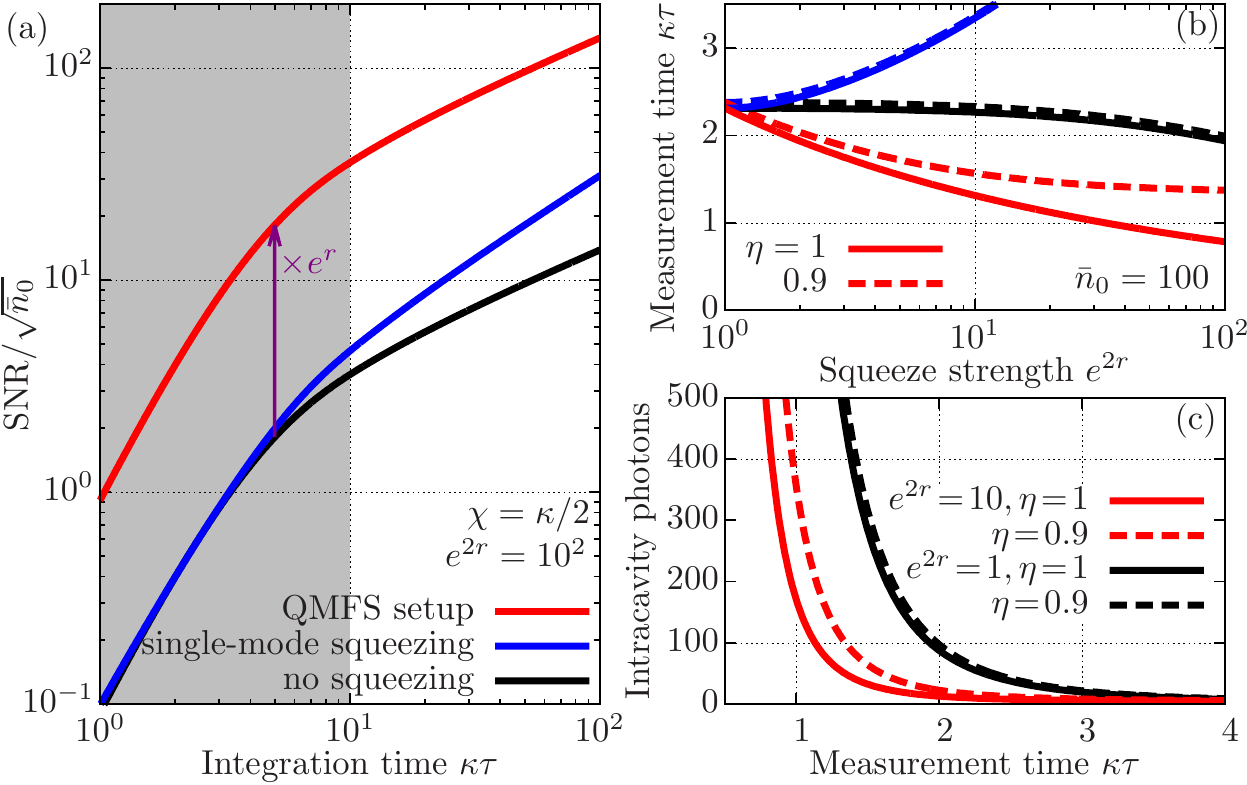}
\caption{
(a) SNR as a function of integration time $\tau$ for different protocols: 
coherent state drive (black), displaced single-mode squeeze state (blue), two-mode squeezed QMFS setup (red).
We assume an optimal dispersive shift $\chi=\kappa/2$;
for the QMFS setup the cavity is presqueezed ($t_0 \ll -1 / \kappa $) with squeeze strength $\Exp{2r}=100$.
The coherent drive is turned on at $\tau = 0$.  
For the single-mode case, for each $\tau$ we optimize the squeeze strength $\Exp{2r}\in[1,100]$ and angle (see~\cite{SM}).
The QMFS scheme gives an exponential SNR enhancement, especially in the most interesting regime where $\tau \sim 1 / \kappa$ (shaded region).
(b) Integration time $\tau$ required to achieve a fidelity $F=99.99\,\%$, as a function of $e^{2r}$; 
parameters as in (a), except that $\no = 100$.  
Black lines correspond to an unsqueezed drive, where the drive strength is increased such that the intracavity photon number is the same as in the QMFS scheme, i.e. $\no \rightarrow \no+4\sinh^2r$.  The solid curves correspond to the case of no photon losses (efficiency $\eta = 1$), while the dashed curves correspond to $\eta = 0.9$.  (c)  Total intracavity photon number needed to achieve $F=99.99\,\%$ in a measurement time $\tau$.  
Even with nonzero photon losses, the use of squeezing can dramatically reduce the number of intracavity photons.}
\label{figSNR}
\end{figure}

Next, consider what happens if we instead inject a displaced squeezed state (squeeze parameter $r$) into the cavity.
As already discussed, 
this is not as beneficial as one would hope, as one always sees the noise of the antisqueezed quadrature ($\propto e^{2r}$)~\cite{Barzanjeh_2014,SM}.  
Consider the optimal case $\phiqb = \pi/2$ which maximizes the signal.
For large $\tau$, the noise behaves as
\begin{align}
\moy{\hat{M}_N^2} & \simeq
\kappa\tau[\sin^2(\theta)\Exp{-2r}+\cos^2(\theta)\Exp{2r}]\nonumber\\
& + 2\sqrt{2} \sinh(2r) \cos(2\theta-3\pi/4),
\label{SNRsinglemode}
\end{align}
where we have dropped terms that decay exponentially with $\kappa \tau$.
The first line of Eq.~\eqref{SNRsinglemode} dominates in the long-time limit, and represents the contribution from zero-frequency noise in the output field.
For this line, the choice $\theta = \pi/2$ 
cancels the contribution from the amplified quadrature, and leads to an exponential reduction in the noise compared to a coherent state drive~\cite{Barzanjeh_2014}.  In contrast, the second line
of Eq.~\eqref{SNRsinglemode} describes the contribution from initial short-time fluctuations; the noise from the antisqueezed quadrature here remains even if $\theta = \pi/2$.  
As a result, increasing $r$ indefinitely does not improve the SNR; for a given $\tau$ there is an optimal value [see Fig.~\ref{figSNR}~(b--c)].  
This then leads to generally modest
enhancement of SNR compared to a simple coherent state drive~\cite{Barzanjeh_2014}; in particular, there is almost no improvement in the most relevant case where 
$\tau\sim1/\kappa$ [shaded region in Fig.~\ref{figSNR}(a)].  
Optimized squeezing leads at best to the scaling $N^{3/4}$ with input photon number, similar to a Mach-Zehnder interferometer driven with squeezed light~\cite{SM,Caves_1981,Giovannetti_2004}.

{\it Negative freqencies and two-mode squeezing.-- }
To avoid having the measurement corrupted by the antisqueezed quadrature, one ideally wants to squeeze {\it both} quadratures of the input light.
While this is impossible with a single cavity, 
it becomes conceivable using joint quadratures of two cavities.  If $\hat{a}_j = (\hat{X}_j + i \hat{Y}_j )/2$ ($j=1,2$) are the annihilation operators for the two cavities (in an interaction
picture with respect to the free cavity Hamiltonians), 
we define $\hat{X}_{\pm} = (\hat{X}_1 \pm \hat{X}_2) / \sqrt{2}$, $\hat{Y}_{\pm} = (\hat{Y}_1 \pm \hat{Y}_2) / \sqrt{2}$.  
Since $X_{-}$ and $Y_{+}$ commute, they can be squeezed simultaneously, resulting in a two-mode squeezed state~\cite{Drummond-Ficek}.
The relevant nonzero input-field noise correlators are 
$\moy{\hat{X}_{\mp}(t)\hat{X}_{\mp}(t')}=\moy{\hat{Y}_{\pm}(t)\hat{Y}_{\pm}(t')}=\Exp{\mp 2r}\delta(t-t')$.
We stress that such states have already been produced in circuit QED~\cite{Flurin_2012,Eichler_2014}.

This squeezing by itself is not enough: we also need the dynamics of these
joint quadratures to mimic the behavior of $\hat{X}$ and $\hat{Y}$ in a single cavity, such that the two qubit states still give rise to a simple rotation of the vector formed by $(X_{-}, Y_{+})$.
Such a dynamics is generated by the simple Hamiltonian~\cite{Wasilewski_2010,Tsang_2010} 
\begin{eqnarray}
	H & = & \tfrac{1}{2}\chi(\hat{X}_+\hat{X}_-+\hat{Y}_+\hat{Y}_-)\hat{\sigma}_z =\chi(\hat{a}_1^\dag\hat{a}_1-\hat{a}_2^\dag\hat{a}_2)\hat{\sigma}_z.
	\label{HXY}
\end{eqnarray}
The qubit thus needs to couple dispersively to both cavities, with equal-magnitude but opposite-signed couplings.
The resulting dynamics is illustrated in Fig.~\ref{figscheme}(d): an incident field with $\langle \hat{Y}_{+} \rangle = 0, \langle \hat{X}_- \rangle \neq 0$ is rotated in a qubit-state dependent manner, resulting
in an output field with $\langle \hat{Y}_{+} \rangle \neq 0$ (i.e.~the measurement signal).
Note that the squeezed quadratures $X_-, Y_+$ are never mixed with the antisqueezed
quadratures $X_+, Y_-$, hence this amplification will not limit our scheme.  We also stress that the two cavities need not have the same 
frequency.

The measurement protocol involves first turning on the vacuum two-mode squeezed drive at a time $t = t_0 \leq 0$, and then turning on the coherent cavity drive(s) at $t=0$.
This coherent drive (which displaces along $X_{-}$ but not $Y_{+}$) could be realized by driving one or both the cavities.
We take the optimal case where both cavities are driven and let $\no\kappa/8$ denote the photon flux incident on each cavity due to the coherent drives.  
The measurement signal in $Y_+$ can be constructed from the quadratures $Y_{j,\out}$ of the output field leaving each cavity.  In what follows, we consider the limit $\kappa t_0\ll -1$, such that the measurement is not corrupted by any initial nonsqueezed vacuum in the cavity~\cite{SM}.

The measurement operator is now $\hat{M}=\sqrt{\kappa}\int_0^\tau\mathrm{d}t\hat{Y}_{+,\out}(t)$.  
As expected, one finds that this output quadrature is always squeezed, and hence the imprecision noise is always described by $\moy{\hat{M}_N^2}=\Exp{-2r}\kappa\tau$, independent of $\chi$.
As desired, the noise is now exponentially reduced with respect to standard dispersive readout, leading to an exponential improvement of SNR, i.e. $\SNR_r(\tau)=\Exp{r}\,\SNR_\alpha(\tau)$ for all integration times $\tau$.  This is in stark contrast to the single-mode approach, where such an enhancement was only possible at extremely long times, $\kappa\tau\gtrsim\Exp{4r}$ [cf.~Eq.~\eqref{SNRsinglemode}].
The SNR is plotted in Fig.~\ref{figSNR}(a) as a function of integration time~$\tau$, with comparisons against the single-mode squeezing and no-squeezing cases; our two-mode scheme realizes dramatic improvements in the most interesting regime where $\tau$ is not much larger than $1/\kappa$.  
The integration time $\tau$ required to achieve a measurement fidelity $F=1-\erfc(\SNR/2)/2$ of $99.99\,\%$ is plotted against squeezing strength in Fig.~\ref{figSNR}(b).
Again, the QMFS scheme results in dramatic improvements.

{\it Heisenberg-limited scaling.-- }
We now show that the SNR scales as the number of photons $N$ used for the measurement rather than its square root $\sqrt{N}$ as it is the case for the standard dispersive readout~\cite{Giovannetti_2004}. For this, we define the temporal mode $\hat{A}=\frac{1}{\sqrt{\tau}}\int_0^\tau\mathrm{d}t[\hat{d}_{\In,1}(t)+\hat{d}_{\In,2}(t)]$~\cite{Yurke_1986} where the operator $\hat{d}_{\In,j}$ describes fluctuations in the resonator-$j$ input field. The total number of input photons $N=N_s+N_d$ has a contribution from squeezing $N_s=\moy{\hat{A}^\dag\hat{A}}=2\sinh^2r$ and $N_d$ from the coherent displacement.  Focusing on times $\tau \gg 1 / \kappa$, we can ignore the transient response to the coherent drive, and hence $N_d =\frac{1}{4}\no\kappa\tau$. Fixing $N$ and taking $t_0 \ll -1/\kappa$, the optimal SNR is obtained for $N_s=N^2/[2(N+1)]$, and is
\begin{equation}
\SNR_\mathrm{opt}=2|\sin\phiqb|N\sqrt{1+2/N} \rightarrow 2|\sin\phiqb|N,
\label{heisenberg}
\end{equation} 
where we have taken the large $N$ limit.  
Eq.~\eqref{heisenberg} corresponds to true Heisenberg scaling for any value of the dispersive coupling.
Such scaling is {\it not} possible using single-mode squeezed input light (see \cite{SM}).

Our QMFS scheme also shows an improved, Heisenberg-like scaling of the SNR with the intracavity photon number $\bar{n}$.  
Note that the SNR for the QMFS scheme 
has the same form as the SNR for a standard ($r=0$) dispersive readout made using a larger drive flux $\no\Exp{2r}$.
If we fix the intra\-cavity photon number $\bar{n}=\no\cos^2(\phiqb/2)+2\sinh^2r$ and optimize~$r$, the resulting SNR scales as 
$\SNR_\mathrm{opt}\simeq2|\sin(\phiqb/2)|\bar{n}\sqrt{\kappa\tau}$, as opposed to the conventional $\SNR_\alpha\propto\sqrt{\bar{n}}$.

{\it Robustness against imperfections.-- }
Our discussion of the QMFS scheme so far has assumed a broadband, pure squeezing source.
The purity of the squeezing is, however, not crucial; our scheme is insensitive to the antisqueezed quadratures, and hence it is not essential that their variances be as small as possible.  
For a finite squeezing bandwidth $\Gamma$, the input 
squeezing spectrum will typically have a Lorentzian line shape~\cite{Walls-Milburn}.
We find that the effects of a finite bandwidth are equivalent to an effective reduction of the squeezing strength; the SNR for the scheme is simply reduced by a prefactor
$\sqrt{\Gamma\tau/[\Gamma\tau+(\Exp{2r}-1)(1-\Exp{-\Gamma\tau})]}$~\cite{SM}.
One thus only needs a modest bandwidth, e.g.\ $\Gamma\sim10\kappa$ is enough for $\kappa\tau\sim10$ and $\Exp{2r}\sim10$.

The lack of any enhanced Purcell decay is also crucial, as in our protocol the squeezing is turned on well before the coherent measurement tone.
Having a finite squeezing bandwidth can in fact be an advantage as it helps suppress Purcell decay of the qubit.  
This decay 
corresponds to relaxation of the qubit by photon emission from the cavity~\cite{houck:2008a}. 
As typical detunings $\Delta \gg \kappa$, there is a wide range of ideal squeezing bandwidths satisfying $\kappa\ll\Gamma\ll\Delta$.  
Such bandwidths are large enough to allow a full enhancement of the SNR (with $\tau \gtrsim 1 / \kappa$), and small enough that the squeezing does not appreciably modify cavity-induced Purcell decay (see~\cite{SM}).  

\begin{figure}
\includegraphics[width=\columnwidth]{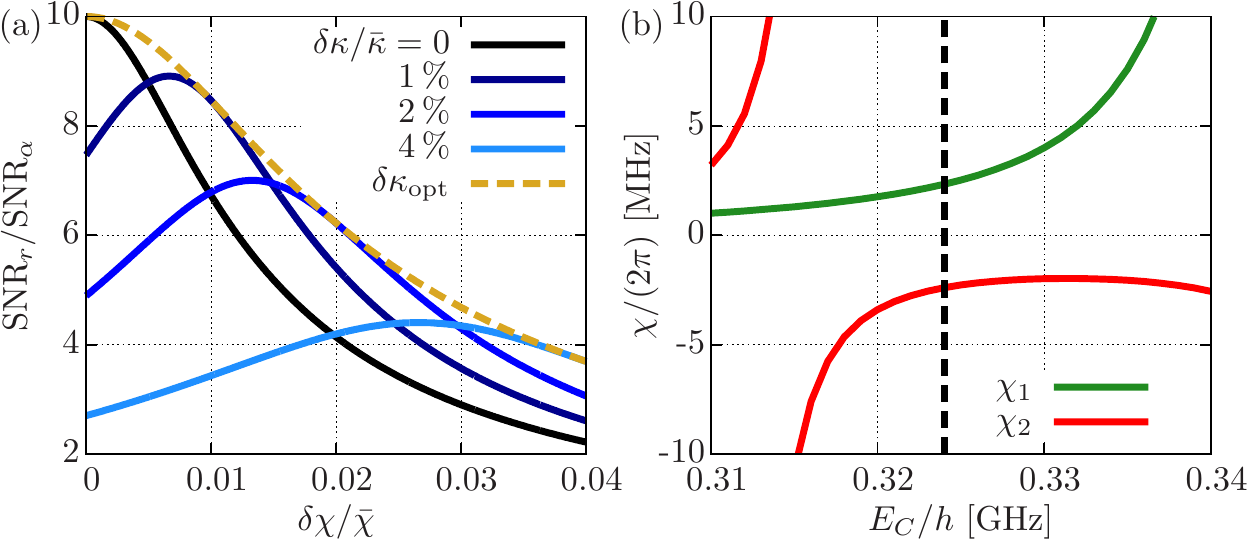}
\caption{
(a) SNR enhancement as a function of the dispersive shift asymmetry ($\chi_{1,2} =\delta\chi \pm\bar{\chi}$) for different resonator linewidth asymmetries ($\kappa_{1,2} = \bar{\kappa} \pm \delta\kappa$) calculated for $\bar{\chi}=\bar{\kappa}/2$, $\kappa\tau=10$ and $\Exp{2r}=100$.
The dashed line is the maximal SNR obtained by optimizing $\delta\kappa$.
(b) Calculated dispersive shifts as a function of transmon anharmonicity $E_C$ from a numerical diagonalization of transmon-resonator system for each of the resonators. 
The parameters are 
$E_J/h = 25\,\mathrm{GHz}$, 
$\omega_1/2\pi = 7.6\,\mathrm{GHz}$, 
$\omega_2/2\pi = 7.9\,\mathrm{GHz}$, 
$g_1/2\pi = 8\,\mathrm{MHz}$ and 
$g_2/2\pi = 15\,\mathrm{MHz}$. 
The vertical dashed line shows a typical value of $E_C$ that leads to equal and opposite dispersive shifts.}
\label{figcQED}
\end{figure}

Another nonideality is asymmetry in the system parameters.  
While the two cavity frequencies can differ, we have assumed so far that they have identical damping rates ($\kappa_1=\kappa_2=\kappa$) and that the dispersive coupling strengths satisfy $\chi_1=-\chi_2 = \chi$.  
Deviation from either of these conditions breaks the symmetry yielding a QMFS, causing an unwanted coupling between the squeezed quadratures $(\hat{X}_{-},\hat{Y}_{+})$ and the antisqueezed quadratures $(\hat{X}_{+},\hat{Y}_{-})$. 
The structure of the QMFS can persist in the presence of asymmetries for long measurement times $\kappa\tau\gg1$,
under the condition~\cite{SM}
\begin{equation}
\frac{\chi_1 + \chi_2 }{\chi_1-\chi_2 }
=\frac{\kappa_1-\kappa_2}{\kappa_1+\kappa_2}.
\label{asym}
\end{equation}
The SNR enhancement can however be preserved for measurement times $\tau\sim1/\kappa$ by optimizing $\delta\kappa/\delta\chi$, 
as illustrated in Fig.~\ref{figcQED}(a).
Although this might not be necessary in practice, all parameters in Eq.~\eqref{asym} can be tuned \textit{in~situ}~\cite{sandberg:2008a,Koch_2007,yin:2013a} thereby greatly relaxing the constraints on the system.

Finally, like any scheme employing squeezing, photon losses effectively replace squeezed fluctuations with ordinary vacuum, causing the 
SNR improvement to saturate as a function of squeezing strength~\cite{SM}.  Despite this, our scheme still yields considerable advantages for finite loss rates, see Figs.~\ref{figSNR}(b) and (c).

{\it Implementation in circuit QED.-- }We now turn to a possible realization of this protocol in circuit QED. All parameters discussed here are readily achievable experimentally. As illustrated in Fig.~\ref{figcircuit}, a transmon qubit is coupled to two resonators, one in the usual dispersive regime ($\Delta > E_C$) while the other in the ``straddling'' regime ($\Delta < E_C$)~\cite{Koch_2007,inomata:2012a}. Here, $\Delta$ is the qubit-resonators detuning and $E_C$ the transmon anharmonicity.  This yields dispersive couplings $\chi$ having opposite signs as required, see Fig.~\ref{figcQED}(b). An alternative strategy is to use a fluxonium or a flux qubit which exhibits a richer dispersive shift profile~\cite{Zhu_2013}. 
Note that either approach does not entail a sacrifice of qubit coherence via enhanced Purcell decay~\cite{SM}.
The displaced two-mode squeezed state required at the input can either be generated by a NDPA such as the Josephson parametric converter~\cite{Flurin_2012}, a Josephson paramp~\cite{Eichler_2011} or the Bose-Hubbard dimer~\cite{Eichler_2014}.

{\it Conclusion.-- }
We have presented a realistic measurement protocol that allows one to exponentially enhance dispersive measurement using two-mode squeezed light, enabling Heisenberg-limited scaling
even with large dispersive couplings.  Our scheme
crucially makes use of a special symmetry in the dynamics of joint cavity quadratures, a so-called ``quantum-mechanics-free subsystem''. 
It could be straightforwardly generalized to allow Heisenberg-limited scaling in any interferometric setup having large signal phase shifts.

{\it Acknowledgements.-- }
We thank Irfan Siddiqi, Michel Devoret, Shabir Barzanjeh, Samuel Boutin, Shruti Puri, Baptiste Royer and Amit Munje for many useful discussions. 
This work was supported by the Army Research Office under Grant W911NF-14-1-0078, INTRIQ, NSERC and CIFAR.

\clearpage


\section*{Supplementary material}

# Supplemental Material for
# "Heisenberg-limited qubit readout with two-mode squeezed light"

Nicolas Didier,[1,2] Archana Kamal,[3] William D. Oliver,[3,4] Alexandre Blais,[2,5] and Aashish A. Clerk[1]

[1] *Department of Physics, McGill University, 3600 rue University, Montreal, Quebec H3A 2T8, Canada*
[2] *Département de Physique, Université de Sherbrooke,
2500 boulevard de l'Université, Sherbrooke, Québec J1K 2R1, Canada*
[3] *Research Laboratory of Electronics, Massachusetts Institute of Technology, Cambridge, Massachusetts 02139, USA*
[4] *MIT Lincoln Laboratory, 244 Wood Street, Lexington, Massachusetts 02420, USA*
[5] *Canadian Institute for Advanced Research, Toronto, Ontario M5G 1Z8, Canada*

## I. QUBIT READOUT WITH QMFS AND TMSS

In this section we detail the derivation of the results presented in the paper for qubit readout in a quantum-mechanics-free subsystem (QMFS) and driven by two-mode squeezed states (TMSS).

### A. Noise

We consider a two-cavity setup where a single qubit is dispersively coupled to both cavities with opposite dispersive shifts, i.e. $\chi_1 = -\chi_2 \equiv \chi$. The cavities can have different frequencies. We suppose they have the same damping rate, i.e. $\kappa_1 = \kappa_2 \equiv \kappa$. The cavities are coupled to zero-temperature environments at times $t < t_0$ and, at times $t \geq t_0$, a TMSS is injected at resonance with the cavities. In the rotating frame of the cavity frequencies, we define four quadratures for the TMSS, $X_\pm, Y_\pm$, characterized by the correlations

$$\langle \hat{\mathbf{Z}}_{\text{in}}(t) \hat{\mathbf{Z}}_{\text{in}}^T(t') \rangle = \begin{pmatrix} 1 & 0 & 0 & i \\ 0 & 1 & -i & 0 \\ \hline 0 & i & 1 & 0 \\ -i & 0 & 0 & 1 \end{pmatrix} \Theta(t_0 - t)\delta(t - t') + \begin{pmatrix} e^{-2r} & 0 & 0 & i \\ 0 & e^{-2r} & -i & 0 \\ \hline 0 & i & e^{2r} & 0 \\ -i & 0 & 0 & e^{2r} \end{pmatrix} \Theta(t - t_0)\delta(t - t'), \quad \text{(S1)}$$

where $\hat{\mathbf{Z}} = (\hat{X}_- \ \hat{Y}_+ \ \hat{X}_+ \ \hat{Y}_-)^T$ and $\Theta$ is the unit step function. From the last term of Eq. (S1), we see that the TMSS is composed of two coupled QMSS: the squeezed subsystem $\hat{\mathbf{Z}}_S = (\hat{X}_- \ \hat{Y}_+)^T$ and the anti-squeezed subsystem $\hat{\mathbf{Z}}_A = (\hat{X}_+ \ \hat{Y}_-)^T$. In order to involve only the squeezed QMFS in the dynamics, the Hamiltonian of the system is chosen to be $\hat{H} = \frac{1}{2}\chi(\hat{X}_+\hat{X}_- + \hat{Y}_+\hat{Y}_-)\hat{\sigma}_z$. As explained in the paper, this is equivalent to having opposite dispersive shifts. The Langevin equations for the intracavity quadratures are found from input-output theory [S1]

$$\dot{\hat{\mathbf{Z}}}(t) = \begin{pmatrix} -\frac{1}{2}\kappa & \chi\hat{\sigma}_z & 0 & 0 \\ -\chi\hat{\sigma}_z & -\frac{1}{2}\kappa & 0 & 0 \\ \hline 0 & 0 & -\frac{1}{2}\kappa & \chi\hat{\sigma}_z \\ 0 & 0 & -\chi\hat{\sigma}_z & -\frac{1}{2}\kappa \end{pmatrix} \hat{\mathbf{Z}}(t) - \sqrt{\kappa}\hat{\mathbf{Z}}_{\text{in}}(t). \quad \text{(S2)}$$

The squeezed and anti-squeezed subsystems are indeed clearly decoupled. The output quadratures are calculated from integrating Eq. (S2) and using the input-output boundary condition $\hat{\mathbf{Z}}_{\text{out}} = \sqrt{\kappa}\hat{\mathbf{Z}} + \hat{\mathbf{Z}}_{\text{in}}$ [S1]. The two-time output correlations of the squeezed QMFS are then equal to

$$\langle \hat{\mathbf{Z}}_{S,\text{out}}(t) \hat{\mathbf{Z}}_{S,\text{out}}^T(t') \rangle = e^{-2r}\delta(t-t')\mathbf{1} + \kappa(1-e^{-2r})e^{-\frac{1}{2}\kappa(t+t'-2t_0)}\mathbf{R}[\chi\hat{\sigma}_z(t-t')], \quad \text{(S3)}$$

where $\mathbf{1}$ is the $2\times 2$ identity matrix and $\mathbf{R}[\phi] = \begin{pmatrix} \cos\phi & \sin\phi \\ -\sin\phi & \cos\phi \end{pmatrix}$ the rotation matrix. Here $t, t' > t_0$ and $t_0 \leq 0$. The correlations of the anti-squeezed QMFS are found by replacing $e^{-2r}$ by $e^{2r}$ in Eq. (S3). The noise associated with the measurement operator $\hat{M}(\tau) = \sqrt{\kappa}\int_0^\tau dt \hat{Y}_{\text{out},+}(t)$ then reads

$$\langle \hat{M}^2(\tau) \rangle = e^{-2r}\kappa\tau + 8\cos^2(\tfrac{1}{2}\varphi_{\text{qb}})(1-e^{-2r})[\cosh(\tfrac{1}{2}\kappa\tau) - \cos(\chi\tau)]e^{-\frac{1}{2}\kappa(\tau-2t_0)}. \quad \text{(S4)}$$

The measurement noise is identical for the two qubit states, $\langle \hat{\sigma}_z \rangle = \pm 1$, for all times $\tau$ and $t_0$. The last terms in Eqs. (S3)–(S4) are due to the leakage of the initial vacuum fluctuations. As expected, the anti-squeezed quadratures do not contribute to the measurement noise. When the cavities are sufficiently pre-squeezed, that is when the TMSS is injected well before the measurement ($t_0 \ll -1/\kappa$), the noise is exponentially reduced.

### B. Asymmetries

To understand the impact of realistic asymmetries in the dispersive shifts, i.e. $\chi_2 \neq -\chi_1$, and damping rates, i.e. $\kappa_2 \neq \kappa_1$, we define $\bar{\chi} = (\chi_1 - \chi_2)/2$, $\bar{\kappa} = (\kappa_1 + \kappa_2)/2$, $\delta\chi = (\chi_1 + \chi_2)/2$, $\delta\kappa = (\kappa_1 - \kappa_2)/2$. In the quadrature basis, the Hamiltonian $\hat{H} = (\chi_1 \hat{a}_1^\dagger \hat{a}_1 + \chi_2 \hat{a}_2^\dagger \hat{a}_2)\hat{\sigma}_z$ reads

$$\hat{H} = \tfrac{1}{2}\bar{\chi}(\hat{X}_+\hat{X}_- + \hat{Y}_+\hat{Y}_-)\hat{\sigma}_z + \tfrac{1}{4}\delta\chi(\hat{X}_+^2 + \hat{Y}_+^2 + \hat{X}_-^2 + \hat{Y}_-^2)\hat{\sigma}_z. \tag{S5}$$

Including the asymmetries, the Langevin equations are

$$\dot{\hat{\mathbf{Z}}}(t) = \begin{pmatrix} -\tfrac{1}{2}\bar{\kappa} & \bar{\chi}\hat{\sigma}_z & -\tfrac{1}{2}\delta\kappa & \delta\chi\hat{\sigma}_z \\ -\bar{\chi}\hat{\sigma}_z & -\tfrac{1}{2}\bar{\kappa} & -\delta\chi\hat{\sigma}_z & -\tfrac{1}{2}\delta\kappa \\ -\tfrac{1}{2}\delta\kappa & \delta\chi\hat{\sigma}_z & -\tfrac{1}{2}\bar{\kappa} & \bar{\chi}\hat{\sigma}_z \\ -\delta\chi\hat{\sigma}_z & -\tfrac{1}{2}\delta\kappa & -\bar{\chi}\hat{\sigma}_z & -\tfrac{1}{2}\bar{\kappa} \end{pmatrix} \hat{\mathbf{Z}}(t) - \sqrt{\kappa}\hat{\mathbf{Z}}_{\text{in}}(t). \tag{S6}$$

The resulting expression for the noise in the measurement operator $\hat{M}(\tau) = \sqrt{\bar{\kappa}} \int_0^\tau dt \hat{Y}_{\text{out},+}(t)$ is quite cumbersome. One can check that the measurement noise does not depend on the qubit state. The leading term for $\tau \gg 1/\kappa$ is

$$\langle \hat{M}^2(\tau) \rangle = \left\{ 1 + \frac{(\bar{\chi}\delta\kappa - \bar{\kappa}\delta\chi)^2(e^{4r} - 1)}{[(\bar{\chi} + \delta\chi)^2 + \tfrac{1}{4}(\bar{\kappa} + \delta\kappa)^2][(\bar{\chi} - \delta\chi)^2 + \tfrac{1}{4}(\bar{\kappa} - \delta\kappa)^2]} \right\} e^{-2r}\bar{\kappa}\tau. \tag{S7}$$

To eliminate the contribution from the anti-squeezed QMFS, the asymmetries have to be tuned such that $\frac{\delta\kappa}{\bar{\kappa}} = \frac{\delta\chi}{\bar{\chi}}$. Under this condition, the full expression for the noise is

$$\langle \hat{M}^2(\tau) \rangle = e^{-2r}\bar{\kappa}\tau + \frac{2\bar{\kappa}^2 \delta\chi^2}{(\bar{\chi}^2 - \delta\chi^2)(\tfrac{1}{4}\bar{\kappa}^2 + \delta\chi^2)} \sinh 2r$$
$$+ \frac{\bar{\kappa}^2 \delta\chi}{(\bar{\chi}^2 + \tfrac{1}{4}\bar{\kappa}^2)(\tfrac{1}{4}\bar{\kappa}^2 + \delta\chi^2)(\bar{\chi} + \delta\chi)} \left\{ \left[\tfrac{1}{4}\bar{\kappa}^2 - \bar{\chi}\delta\chi\right] \cos(\chi_1 \tau) - \tfrac{1}{2}\bar{\kappa}\left[\bar{\chi} + \delta\chi\right]\sin(\chi_1 \tau) \right\} e^{-\tfrac{1}{2}\kappa_1 \tau} \sinh 2r$$
$$+ \frac{\bar{\kappa}^2 \delta\chi}{(\bar{\chi}^2 + \tfrac{1}{4}\bar{\kappa}^2)(\tfrac{1}{4}\bar{\kappa}^2 + \delta\chi^2)(\bar{\chi} - \delta\chi)} \left\{ \left[\tfrac{1}{4}\bar{\kappa}^2 + \bar{\chi}\delta\chi\right] \cos(\chi_2 \tau) + \tfrac{1}{2}\bar{\kappa}\left[\bar{\chi} - \delta\chi\right]\sin(\chi_2 \tau) \right\} e^{-\tfrac{1}{2}\kappa_2 \tau} \sinh 2r. \tag{S8}$$

The remaining terms in $\langle \hat{M}^2(\tau) \rangle$ are constant or decaying exponentially. For finite integration times $\tau \sim 1/\kappa$, the SNR can be maximized numerically by optimizing $\delta\chi$ and $\delta\kappa$.

### C. Squeezing bandwidth

For a finite squeezing bandwidth $\Gamma$, the input field correlations are

$$\langle \hat{\mathbf{Z}}_{\text{in}}(t) \hat{\mathbf{Z}}_{\text{in}}^T(t') \rangle = \begin{pmatrix} 1 & 0 & 0 & i \\ 0 & 1 & -i & 0 \\ 0 & i & 1 & 0 \\ -i & 0 & 0 & 1 \end{pmatrix} \delta(t - t') + \begin{pmatrix} e^{-2r} - 1 & 0 & 0 & 0 \\ 0 & e^{-2r} - 1 & 0 & 0 \\ 0 & 0 & e^{2r} - 1 & 0 \\ 0 & 0 & 0 & e^{2r} - 1 \end{pmatrix} \frac{\Gamma}{2} e^{-\Gamma|t-t'|} \Theta(t - t_0) \Theta(t' - t_0), \tag{S9}$$

The noise then reads

$$\langle \hat{M}^2(\tau) \rangle = \frac{\Gamma\tau + (e^{2r}-1)(1-e^{-\Gamma\tau})}{\Gamma\tau} e^{-2r}\kappa\tau$$

$$+ 8\left(1 - e^{-2r}\right)\cos^2(\tfrac{1}{2}\varphi_{\rm qb})\frac{\Gamma(\Gamma - \tfrac{1}{2}\kappa)}{\chi^2 + (\Gamma - \tfrac{1}{2}\kappa)^2}\left[\cosh(\tfrac{1}{2}\kappa\tau) - \cos(\chi\tau)\right]e^{-\tfrac{1}{2}\kappa(\tau - 2\kappa t_0)}$$

$$- 4\left(1 - e^{-2r}\right)\cos^2(\tfrac{1}{2}\varphi_{\rm qb})\frac{1 - e^{-\Gamma\tau}}{\chi^2 + (\Gamma - \tfrac{1}{2}\kappa)^2}\left\{\left[\chi^2 - \tfrac{1}{2}\kappa(\Gamma - \tfrac{1}{2}\kappa)\right]\left[\cos(\chi(\tau - t_0))e^{-\tfrac{1}{2}\kappa\tau} - \cos(\chi t_0)\right]\right.$$

$$\left. + \Gamma\chi\left[\sin(\chi(\tau-t_0))e^{-\tfrac{1}{2}\kappa\tau} + \sin(\chi t_0)\right]\right\} e^{(\Gamma + \tfrac{1}{2}\kappa)t_0}. \quad (S10)$$

The infinite bandwidth result Eq. (S4) is recovered in the limit $\Gamma\tau \to \infty$.

## II. QUBIT READOUT WITH SINGLE-MODE SQUEEZED STATES (SMSS)

### A. Noise

We consider a cavity dispersively coupled to a qubit. As described for the QMFS protocol, the cavity is coupled to a zero-temperature environment at times $t < t_0$ and, at times $t \geq t_0$, a SMSS is injected at resonance with the cavity. We note $\theta$ the squeeze angle. In the rotating frame of the cavity frequency, we define two quadratures for the SMSS, $X, Y$, characterized by the correlations

$$\langle \hat{\mathbf{Z}}_{\rm in}(t)\hat{\mathbf{Z}}_{\rm in}^T(t')\rangle = \begin{pmatrix} 1 & -i \\ i & 1 \end{pmatrix}\Theta(t_0-t)\delta(t-t') + \begin{pmatrix} e^{2r}\cos^2\theta + e^{-2r}\sin^2\theta & -i + \sin 2\theta \sinh 2r \\ i + \sin 2\theta \sinh 2r & e^{-2r}\cos^2\theta + e^{2r}\sin^2\theta \end{pmatrix}\Theta(t-t_0)\delta(t-t'), \quad (S11)$$

where $\hat{\mathbf{Z}} = (\hat{X}\ \hat{Y})^T$. The Langevin equations for the intracavity quadratures are

$$\dot{\hat{\mathbf{Z}}}(t) = \begin{pmatrix} -\tfrac{1}{2}\kappa & \chi\hat{\sigma}_z \\ -\chi\hat{\sigma}_z & -\tfrac{1}{2}\kappa \end{pmatrix}\hat{\mathbf{Z}}(t) - \sqrt{\kappa}\hat{\mathbf{Z}}_{\rm in}(t). \quad (S12)$$

In the Fourier space and in the frame of the homodyne angle $\varphi_{\rm h}$, the output quadratures $\hat{\mathbf{Z}}_{{\rm out},\varphi_{\rm h}} = \mathbf{R}[\varphi_{\rm h}]\hat{\mathbf{Z}}_{\rm out}$ are

$$\hat{\mathbf{Z}}_{{\rm out},\varphi_{\rm h}}(\omega) = -\exp\left\{i\arctan\frac{\kappa\omega}{\tfrac{1}{4}\kappa^2 + \chi^2 - \omega^2}\right\}\mathbf{R}\left[\varphi_{\rm h} - \theta + \arctan\frac{\kappa\chi\hat{\sigma}_z}{\tfrac{1}{4}\kappa^2 - \chi^2 + \omega^2}\right]\hat{\mathbf{Z}}_{\rm in}(\omega). \quad (S13)$$

The rotation angle is frequency-dependent, the squeezed and anti-squeezed quadratures are thus simultaneously involved in the quadratures of the output field for any value of $\varphi_{\rm h}$. For the qubit state $|1\rangle$, the noise of the measurement operator $\hat{M}_N(\tau) = \sqrt{\kappa}\int_0^\tau dt\, \hat{Y}_{{\rm out},\varphi_{\rm h}}(t)$ is then equal to

$$\langle \hat{M}^2_{N,|1\rangle}(\tau)\rangle = \kappa\tau\left\{\cos^2(\theta - \varphi_{\rm h} - \varphi_{\rm qb})e^{-2r} + \sin^2(\theta - \varphi_{\rm h} - \varphi_{\rm qb})e^{2r}\right\}$$

$$+ 4\sinh 2r \sin\varphi_{\rm qb}\cos\tfrac{1}{2}\varphi_{\rm qb}\left\{\sin[2(\theta-\varphi_{\rm h}) - \tfrac{3}{2}\varphi_{\rm qb}] - e^{-\tfrac{1}{2}\kappa\tau}\sin[2(\theta-\varphi_{\rm h}) - \tfrac{3}{2}\varphi_{\rm qb} - \chi\tau]\right\}$$

$$+ 16\sinh^2 r \cos^2\tfrac{1}{2}\varphi_{\rm qb}\left[\cos\chi\tau - \cosh\tfrac{1}{2}\kappa\tau\right]e^{-\tfrac{1}{2}\kappa(\tau - 2t_0)}$$

$$+ 8\sinh 2r \cos^3\tfrac{1}{2}\varphi_{\rm qb}\cos[2(\theta-\varphi_{\rm h}) - \tfrac{3}{2}\varphi_{\rm qb} - \chi(\tau - 2t_0)](1 - \cos\chi\tau\cosh\tfrac{1}{2}\kappa\tau)e^{-\tfrac{1}{2}\kappa(\tau - 2t_0)}$$

$$- 8\sinh 2r \cos^3\tfrac{1}{2}\varphi_{\rm qb}\sin[2(\theta-\varphi_{\rm h}) - \tfrac{3}{2}\varphi_{\rm qb} - \chi(\tau - 2t_0)]\sin\chi\tau\sinh\tfrac{1}{2}\kappa\tau e^{-\tfrac{1}{2}\kappa(\tau - 2t_0)}. \quad (S14)$$

The noise $\langle \hat{M}^2_{N,|0\rangle}(\tau)\rangle$ for the qubit state $|0\rangle$ is obtained by replacing $\varphi_{\rm qb} \to -\varphi_{\rm qb}$ and $\chi \to -\chi$ in Eq. (S14). The anti-squeezed quadrature cannot be eliminated in the measurement noise. As a consequence, for a given measurement time $\tau$, there is an optimal value of the squeeze parameter $r$ that minimizes the measurement noise.

### B. Optimal SNR

To calculate the SNR, defined by ${\rm SNR} = \left|M_{S,|0\rangle} - M_{S,|1\rangle}\right|/(\langle \hat{M}^2_{N,|0\rangle}\rangle + \langle \hat{M}^2_{N,|1\rangle}\rangle)^{1/2}$, the signal can be readily calculated from the Langevin equation Eq. (S12), giving

$$M_{S,|0\rangle} - M_{S,|1\rangle} = 2\sqrt{\bar{n}_0}\cos(\varphi_{\rm h})\sin(\varphi_{\rm qb})\kappa\tau[1 - F(\tau)], \quad F(\tau) = \frac{4}{\kappa\tau}\frac{\chi}{\kappa}\left[\sin(\varphi_{\rm qb}) - \sin(\varphi_{\rm qb} + \chi\tau)e^{-\tfrac{1}{2}\kappa\tau}\right]. \quad (S15)$$

The noise contribution is calculated from Eq. (S14), we take the limit $t_0 \to -\infty$ for simplicity,

$$\langle \hat{M}^2_{N,|0\rangle} \rangle + \langle \hat{M}^2_{N,|1\rangle} \rangle = 2\kappa\tau \left\{ \cosh(2r) - \cos[2(\varphi_h - \theta)] \sinh(2r) G(\tau) \right\}, \tag{S16}$$

$$G(\tau) = \cos(2\varphi_{\rm qb}) + \frac{4}{\kappa\tau} \sin(\varphi_{\rm qb}) \cos(\tfrac{1}{2}\varphi_{\rm qb}) \left[ \sin(\tfrac{3}{2}\varphi_{\rm qb}) - \sin(\tfrac{3}{2}\varphi_{\rm qb} + \chi\tau) e^{-\tfrac{1}{2}\kappa\tau} \right]. \tag{S17}$$

The corresponding SNR is optimized with $\varphi_h = 0$; $\theta = 0$ if $G(\tau) > 0$ and $\theta = \tfrac{\pi}{2}$ if $G(\tau) < 0$; $r_{\rm opt} = \tfrac{1}{2}\operatorname{arctanh}|G(\tau)|$,

$$\text{SNR}_{\rm SMSS}(r_{\rm opt}) = \sqrt{2\bar{n}_0 \kappa \tau} |\sin(\varphi_{\rm qb})| \frac{|1 - F(\tau)|}{[1 - G^2(\tau)]^{1/4}} = \sqrt{2\bar{n}_0 \kappa \tau} |\sin(\varphi_{\rm qb})| e^{r_{\rm opt}} \frac{|1 - F(\tau)|}{[1 + |G(\tau)|]^{1/2}}. \tag{S18}$$

The optimal squeeze strength is rather small, leading to a modest improvement of the SNR. For instance, for $\chi = \tfrac{1}{2}\kappa$: at $\tau = 1/\kappa$, $e^{2r_{\rm opt}} \simeq 1.8$ and $\text{SNR}_{\rm SMSS}(r_{\rm opt})/\text{SNR}_{\rm SMSS}(0) \simeq 1.1$; at $\tau = 10/\kappa$, $e^{2r_{\rm opt}} \simeq 3.0$ and $\text{SNR}_{\rm SMSS}(r_{\rm opt})/\text{SNR}_{\rm SMSS}(0) \simeq 1.3$; at $\tau = 100/\kappa$, $e^{2r_{\rm opt}} \simeq 9.9$ and $\text{SNR}_{\rm SMSS}(r_{\rm opt})/\text{SNR}_{\rm SMSS}(0) \simeq 2.2$.

### C. Scaling

We now focus on the situation that maximizes the signal, i.e. $\chi = \kappa/2$. For $\kappa t_0 \to -\infty$ and neglecting the transient dynamics due to the drive as well as the terms exponentially decaying with $\kappa\tau$, the SNR reads

$$\text{SNR}^2_{\rm SMSS} = \frac{32 N_s (N - N_s)}{1 + \frac{16 N_s^2}{\kappa\tau}} = \frac{32 N_s (N - N_s)^2}{N - N_s + 4\bar{n}_0 N_s^2}, \tag{S19}$$

where we used $N_s = \sinh^2 r \simeq \tfrac{1}{4} e^{2r}$ for large squeeze parameters and $N = N_d + N_s$ with $N_d = \bar{n}_0 \kappa \tau / 4$. Unlike the QMFS scheme, the SNR cannot be written only in terms of $N$ and $N_s$; the parameter $\kappa\tau$ or $\bar{n}_0$ is also involved.

For a fixed drive strength $\bar{n}_0$, we calculate the optimal value of $N_s$. The SNR then scales as

$$\text{SNR}_{\rm SMSS,opt} \simeq 8 N^{3/4}/\bar{n}_0^{1/4}, \qquad N_{s,\rm opt} = \tfrac{1}{2}\sqrt{N/\bar{n}_0}, \tag{S20}$$

where we have taken the large $N$ limit. Such a scaling, SNR $\propto N^{3/4}$, is obtained in a Mach-Zehnder interferometer driven with squeezed light [S2, S3].

For a fixed integration time $\kappa\tau$, we calculate the optimal value of $N_s$. The SNR then scales as

$$\text{SNR}_{\rm SMSS,opt} = \sqrt{\sqrt{(16 N^2 + \kappa\tau)\kappa\tau} - \kappa\tau} \simeq 2\sqrt{N}(\kappa\tau)^{1/4}, \quad N_{s,\rm opt} = \frac{\sqrt{16 N^2 \kappa\tau + (\kappa\tau)^2} - \kappa\tau}{16 N} \simeq \tfrac{1}{4}\sqrt{\kappa\tau}, \tag{S21}$$

where we have taken the large $N$ limit. Such a scaling, SNR $\propto N^{1/2}$, is obtained with a coherent drive without squeezed light [S3].

## III. PURCELL DECAY

### A. Purcell decay due to cavity squeezing

The coupling Hamiltonian between a cavity mode and the environment is $H_{\rm diss} = \int_0^\infty d\omega \sqrt{\kappa(\omega)} [\hat{b}^\dagger(\omega) + \hat{b}(\omega)][\hat{a}^\dagger + \hat{a}]$, where $\hat{a}$ is the annihilation operator of the cavity and $\hat{b}(\omega)$ the annihilation operator of the bath at $\omega$. We apply the dispersive transformation to diagonalize the Jaynes-Cummings interaction between the cavity and the qubit in the dispersive regime, where the detuning $\Delta = \omega_a - \omega_1$ between the qubit and the cavity is much larger than their coupling strength $g$. Under this unitary transformation (noted with the superscript $D$) and at lowest order in $g/\Delta$, the cavity operator acquires a contribution from the qubit: $\hat{a}^D = \hat{a} + \tfrac{g}{\Delta}\hat{\sigma}_-$. In the dressed basis, the qubit becomes coupled to the environment of the cavity. Following Ref. [S4], we apply the rotating-wave approximation in the interaction picture to obtain the coupling Hamiltonian between the system and the environment in the dressed basis

$$H^D_{\rm diss} = \int_0^\infty d\omega \sqrt{\kappa(\omega)}\, \hat{b}^\dagger(\omega) \hat{a}\, e^{i(\omega - \omega_1)t} + \frac{g}{\Delta} \int_0^\infty d\omega \sqrt{\kappa(\omega)}\, \hat{b}^\dagger(\omega) \hat{\sigma}_-\, e^{i(\omega - \omega_a)t} + \text{H.c.} \tag{S22}$$

The last term of the coupling Hamiltonian Eq. (S22) results in a decay channel for the qubit. The corresponding damping rate is $\gamma_\kappa = \left(\frac{g}{\Delta}\right)^2 \kappa(\omega_a)$. For a squeezed environment of squeeze parameter $r$ at resonance with the cavity frequency $\omega_1$ and with a bandwidth $\Gamma$, the bath correlators are

$$\langle \hat{b}^\dagger(\omega)\hat{b}(\omega')\rangle = \sinh^2 r \frac{\Gamma^2}{(\omega-\omega_1)^2+\Gamma^2}\delta(\omega-\omega'), \quad \langle \hat{b}(\omega)\hat{b}(\omega')\rangle = \tfrac{1}{2}\sinh(2r)e^{2i\theta}\frac{\Gamma^2}{(\omega-\omega_1)^2+\Gamma^2}\delta(\omega+\omega'-2\omega_1). \quad \text{(S23)}$$

Because the qubit is strongly detuned with the cavity, the squeezing terms will only have a negligible effect on the qubit. The qubit is thus coupled to a thermal bath with the population $\bar{n}_a = \sinh^2 r \frac{\Gamma^2}{\Delta^2+\Gamma^2}$ and the damping rate $\gamma_\kappa$. One can then compute the dynamics of $\langle \hat{\sigma}_z \rangle$

$$\langle \hat{\sigma}_z \rangle(t) = \left[\langle \hat{\sigma}_z \rangle(0) + \frac{1}{2\bar{n}_a+1}\right]e^{-\gamma_s t} - \frac{1}{2\bar{n}_a+1}, \qquad \gamma_s = \gamma_\kappa(2\bar{n}_a+1). \qquad \text{(S24)}$$

The qubit excited state relaxes with the Purcell decay rate $\gamma_s$.

As expected, the contribution from $\bar{n}_a$ to the decay is therefore largely suppressed for $\Delta \gg \Gamma$ and hence pre-squeezing (i.e. $t_0 \neq 0$) does not affect the qubit relaxation rate for $\Gamma \ll \Delta$.

### B. Purcell decay in the straddling regime

To obtain the opposite-signed dispersive shifts required for the TMSS protocol, one of the readout cavities is parked in the straddling regime with the qubit, i.e. $\Delta < 2\pi|f_{ge} - f_{ef}|$, where $f_{ge}$, $f_{ef}$ denote the lowest two qubit transition frequencies. The qubit-resonator coupling strength, however, can be chosen to be proportionately weaker: $g/2\pi \approx$ 10-20 MHz instead of the usual 100-200 MHz used in the standard dispersive regime. As a result, the smaller value of $\Delta$ is compensated by the smaller value of $g$, and the Purcell decay rate is no larger than in the usual case (where both $g$ and $\Delta$ are larger). At the same time, due to the enhanced contributions from the second excited state, a large dispersive shift can still be realized to match the shift induced by the dispersively coupling cavity. Thus, due to the value of $g/\Delta \approx 0.1$ being similar for both the readout cavities, both will have similar Purcell decay rates.

For optimal readout conditions, i.e. $\kappa = 2\chi$, and assuming the squeezing bandwidth to be sufficiently small ($\Gamma/\Delta < 0.1$) so that any decay due to extra photons generated by squeezing can be ignored, we estimate the Purcell decay for the configuration presented in Figure 4(b) of the main text. For $E_J/h = 25$ GHz, $E_C/h = 0.325$ GHz, $g_1/2\pi = 8$ MHz, $g_2/2\pi = 15$ MHz, we obtain a (single mode) Purcell decay time of 18 $\mu$s (for Cavity 1 at 7.6 GHz) and 11 $\mu$s (for Cavity 2 at 7.9 GHz), corresponding to (calculated) qubit frequencies $f_{ge} = 7.72$ GHz, $f_{ef} = 7.35$ GHz. Given the short measurement times [c.f. Figure 3(a) of main text], there should be no appreciable qubit decay, and concomitant loss of readout fidelity, during the measurement.

We note that it is also possible to implement the straddling regime with strongly anharmonic qubits (such as the flux qubits or fluxonium) having higher anharmonicities of few GHz. This allows greater leeway with respect to qubit-resonator detunings and dispersive shift profiles. This was exploited in the experiment reported in [S5] where the straddling regime was realized with a flux qubit. Despite the very large dispersive shifts ($\sim 80$ MHz), the authors of that paper have reported no degradation of the qubit relaxation time due to resonator-induced decay.

### IV. EFFECT OF LOSSES

### A. Photon loss

We first study the effects of photon losses from the cavity output fields by adding a beam splitter at the output of each cavity $j = 1, 2$; these beam splitters combine the cavity output field $\hat{a}_{\text{out},j}$ (which is squeezed) with vacuum noise (operator $\hat{c}_{\text{in},j}$). Letting $\eta$ denote the transmittance, the output field $\hat{\mathcal{A}}_{\text{out},j}$ of each beam splitter is $\hat{\mathcal{A}}_{\text{out},j} = \sqrt{\eta}\hat{a}_{\text{out},j} + \sqrt{1-\eta}\hat{c}_{\text{in},j}$. The output field $\hat{\mathcal{Y}}_{\text{out},+}$ leaving the beam splitters then reads

$$\hat{\mathcal{Y}}_{\text{out},+} = \sqrt{\eta}\,\hat{Y}_{\text{out},+} + \sqrt{1-\eta}\,\hat{y}_{\text{in},+}, \qquad \text{(S25)}$$

where $\hat{y}_{\text{in},+} = \frac{i}{\sqrt{2}}(\hat{c}^\dagger_{\text{in},1} - \hat{c}_{\text{in},1} + \hat{c}^\dagger_{\text{in},2} - \hat{c}_{\text{in},2})$ is the joint quadrature of the vacuum fluctuations that has zero mean and correlations $\langle \hat{y}_{\text{in},+}(t)\hat{y}_{\text{in},+}(t')\rangle = \delta(t-t')$. The vacuum correlations of the joint quadrature actually correspond

to the output signal correlations without squeezing. The measurement operator after the beam splitters is defined as $\hat{M}_\eta(\tau) = \sqrt{\kappa} \int_0^\tau dt\, \hat{\mathcal{Y}}_{\text{out},+}(t)$. From the mean and two-time correlations, the measurement signal and noise read

$$M_{\eta,S}(\tau) = \sqrt{\eta}\hat{M}(\tau), \tag{S26}$$

$$\langle \hat{M}^2_{\eta,N}(\tau)\rangle = \eta\langle \hat{M}^2_N(\tau)\rangle + (1-\eta)\langle \hat{M}^2_N(\tau)\rangle_{r=0}. \tag{S27}$$

The SNR after the beam splitters is finally equal to

$$\text{SNR}_\eta(r) = \left[\frac{1}{\text{SNR}^2(r)} + \frac{1-\eta}{\eta}\frac{1}{\text{SNR}^2(r=0)}\right]^{-\frac{1}{2}}. \tag{S28}$$

The general result Eq. (S28) shows that to fully exploit the properties of squeezed states in quantum metrology, photon loss has to be made minimal; as mentioned in the main text, this is true for any scheme which attempts to exploit squeezing. As the SNR is exponentially enhanced in the absence of photon loss, $\text{SNR}(r) = e^r \text{SNR}(r=0)$, we get

$$\text{SNR}_\eta(r) = \frac{\text{SNR}(r)}{\sqrt{1 + \frac{1-\eta}{\eta}e^{2r}}}. \tag{S29}$$

For non-zero photon loss, the SNR saturates as a function of $r$ (as opposed to increasing indefinitely). Practically speaking, for $\eta \leq 1$, the maximum useful amount of squeezing corresponds to $e^{2r} \sim \frac{\eta}{1-\eta}$; further increases in $r$ do not significantly improve the SNR.

### B. Internal loss

Photon losses occurring inside the cavities (internal loss) have a qualitatively similar effect to downstream losses. We study the effect of an additional decay channel to a reservoir at zero temperature in each cavity. The damping rate to the two-mode squeezed reservoir is noted $\mu\kappa$ and the damping rate to vacuum is noted $(1-\mu)\kappa$, keeping the total damping rate $\kappa$. Following Sec. I A, we first calculate the Langevin equation of the squeezed QMFS,

$$\dot{\hat{\mathbf{Z}}}_S(t) = \begin{pmatrix} -\frac{1}{2}\kappa & \chi\hat{\sigma}_z \\ -\chi\hat{\sigma}_z & -\frac{1}{2}\kappa \end{pmatrix} \hat{\mathbf{Z}}_S(t) - \sqrt{\mu\kappa}\,\hat{\mathbf{Z}}_{S,\text{in}}(t) - \sqrt{(1-\mu)\kappa}\,\hat{\mathbf{z}}_{\text{in}}(t), \tag{S30}$$

where $\hat{\mathbf{z}}_{\text{in}}$ are the joint operators of the vacuum baths. The output field is then calculated using the boundary condition $\hat{\mathbf{Z}}_{S,\text{out}} = \sqrt{\mu\kappa}\hat{\mathbf{Z}}_S + \hat{\mathbf{Z}}_{S,\text{in}}$. The measurement signal turns out to be equal to

$$M_{\mu,S}(\tau) = \mu^2 M_S(\tau). \tag{S31}$$

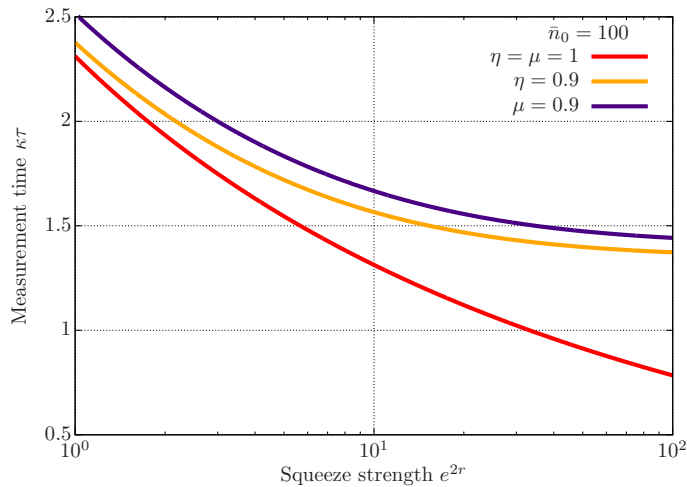

FIG. S1. Measurement time to reach a fidelity of 99.99 % against squeeze strength without loss (red line) and in the presence of photon loss, Eq. (S29) (orange line), and internal loss, Eqs. (S31)–(S32) (indigo line).

Again following Sec. I A, the measurement noise is found to be

$$\langle \hat{M}_N(\tau) \rangle_\mu = e^{-2r} \mu \kappa \tau \left\{ 1 + (e^{2r} - 1)\mu(1-\mu) 4\cos^2(\tfrac{1}{2}\varphi_{\text{qb}}) \left[ 1 - \frac{2}{\kappa \tau} \left( \cos(\varphi_{\text{qb}}) - \cos(\varphi_{\text{qb}} + \chi \tau) e^{-\frac{1}{2}\kappa \tau} \right) \right] \right\}. \quad \text{(S32)}$$

The spurious decay channel reduces the measurement signal and gives rise to a transient behavior to the measurement noise, thereby altering the corresponding $\text{SNR}_\mu$. The effect of the internal loss on the measurement time is plotted in Fig. S1 and compared to the case of photon loss from the cavity output fields.

---



\end{document}